# Application of Quantum Annealing to Training of Deep Neural Networks

Steven H. Adachi[1], Maxwell P. Henderson[2]

**Abstract**

In Deep Learning, a well-known approach for training a Deep Neural Network starts by training a generative Deep Belief Network model, typically using Contrastive Divergence (CD), then fine-tuning the weights using backpropagation or other discriminative techniques.  However, the generative training can be time-consuming due to the slow mixing of Gibbs sampling.  We investigated an alternative approach that estimates model expectations of Restricted Boltzmann Machines using samples from a D-Wave quantum annealing machine. We tested this method on a coarse-grained version of the MNIST data set. In our tests we found that the quantum sampling-based training approach achieves comparable or better accuracy with significantly fewer iterations of generative training than conventional CD-based training.  Further investigation is needed to determine whether similar improvements can be achieved for other data sets, and to what extent these improvements can be attributed to quantum effects.

---

[1] Lockheed Martin Information Systems & Global Solutions, Palo Alto, California, USA
(steven.h.adachi@lmco.com)
[2] Lockheed Martin Information Systems & Global Solutions, King of Prussia, Pennsylvania, USA

# 1 Introduction

In Deep Learning, a well-known approach for training a Deep Neural Network (DNN) starts by training a generative Deep Belief Network (DBN) model, typically using Contrastive Divergence (CD), then fine-tuning the weights using backpropagation or other discriminative techniques. However, the generative training can be time-consuming due to the slow mixing of Gibbs sampling.

One intriguing idea that has been suggested recently is to use sampling from a quantum computer, e.g. a D-Wave quantum annealing machine, in place of classical Gibbs sampling. Theoretically, there is some reason to believe that samples from a quantum computer could yield representative Boltzmann statistics with fewer samples than classical Gibbs sampling. The phenomena of quantum superposition and tunneling imply that certain types of energy landscapes can be more efficiently explored by quantum annealing than classical simulated annealing [1]. The devices made by D-Wave Systems [2] constitute physical implementations of quantum annealing, and there is some evidence for coherent quantum tunneling in a D-Wave device [3]. Similarly, Wiebe et al [4] have shown how gate operations could be used to prepare a coherent analog of the Gibbs state, from which samples could be drawn for training a Boltzmann machine.

In practice, there may be challenges in achieving sufficient fidelity to a Boltzmann machine using a quantum computer, due to the limitations of the quantum computing hardware. Dumoulin et al [5] performed simulations to investigate the feasibility of using the D-Wave machine as a physical implementation of a Restricted Boltzmann Machine (RBM). Their results suggested that the performance of the D-Wave machine as an RBM could be seriously impaired by the limited qubit connectivity, and to a lesser extent by the noise associated with setting parameters on the hardware. Denil and de Freitas [6] also performed simulations of the D-Wave machine acting as a Quantum Boltzmann Machine; however their model only considered implementing hidden units on the quantum processor, and also differed from an RBM in that it allowed connections between hidden units.

Rose [7] implemented a 4/4/4 Deep Boltzmann Machine directly on the D-Wave hardware, using a classical training method previously published by G. Montavon [8], but replacing classical alternating Gibbs sampling with quantum sampling from the D-Wave chip. He found that the quantum approach took fewer iterations of generative training than the classical method to achieve the same Kullback-Leibler divergence score.

We investigated an approach for training DNNs that uses quantum sampling from a D-Wave machine as part of the generative DBN training, and that addresses the previously identified challenges due to the D-Wave's limited qubit connectivity and parameter setting noise. In addition, our approach is also robust enough to accommodate small numbers of faulty qubits. Using this approach we have successfully trained DBNs with up to 32 visible nodes and 32 hidden nodes per RBM layer, on a 512-qubit D-Wave machine with 8 faulty qubits. So far, we have found that the quantum sampling-based training approach achieves comparable or better accuracy with significantly fewer iterations of generative training than conventional CD-based training.

Further investigation is needed to determine whether similar improvements can be achieved for other data sets, and to what extent these improvements can be attributed to quantum effects. As larger quantum annealing machines become available in the future, further investigation is needed to characterize how the performance of the quantum sampling-based training approach scales relative to conventional training as the size of the RBM increases.

In addition, the idea of using a quantum annealing machine to do sampling or inference, as opposed to optimization, could lead to new applications of quantum annealing in addition to Deep Learning. Exact inference on probabilistic graphical models becomes intractable as the number of nodes increases, except when the graph is a tree, and approximate inference methods such as loopy belief propagation are not always guaranteed to converge.[9] Similarly, sampling methods such as Markov Chain Monte Carlo, can have very long mixing times. Quantum sampling (using a quantum annealing machine to draw representative samples from a Boltzmann distribution) could potentially provide an alternative to conventional techniques for sampling and inference in some cases.

## 2   Conventional Approach for Training Deep Belief Networks

This section briefly reviews the well-known approach for generative training of a Deep Belief Network (DBN) using Contrastive Divergence (CD). A DBN is constructed by stacking Restricted Boltzmann Machines (RBMs). An RBM consists of stochastic binary variables arranged into a visible layer and a hidden layer, where inter-layer connections are allowed but intra-layer connections are forbidden. Thus, the connectivity of an RBM can be represented as an undirected bipartite graph.

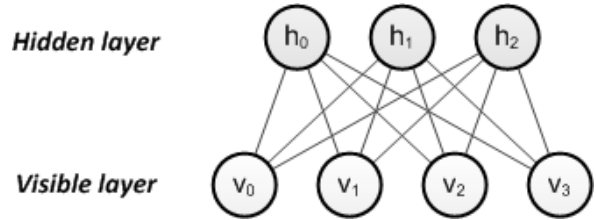

**Figure 1 Restricted Boltzmann Machine (RBM), basic building block of Deep Belief Networks**

The joint probability distribution is defined by a Gibbs distribution

$$P(v, h) = \frac{1}{Z} \exp(-E(v, h)) \qquad (1)$$

with an energy functional

$$E(v, h) = -\sum_{i=1}^{n} b_i v_i - \sum_{j=1}^{m} c_j h_j - \sum_{i=1}^{n} \sum_{j=1}^{m} W_{ij} v_i h_j \qquad v_i, h_j \in \{0,1\} \qquad (2)$$

where $n$ is the number of visible nodes and $m$ is the number of hidden nodes.  The normalization constant

$$Z = \sum_{\{v_k\}} \sum_{\{h_l\}} \exp\left( \sum_k b_k v_l + \sum_l c_l h_l + \sum_{kl} W_{kl} v_k h_l \right) \qquad (3)$$

is known in physics as the partition function.

Because of the bipartite graph structure, the forward and reverse conditional probability distributions for an RBM are both simple sigmoid functions:

$$P(h_j = 1 \mid v) = \text{sigm}\left(c_j + \sum_i W_{ij}\, v_i\right) \tag{4}$$

$$P(v_i = 1 \mid h) = \text{sigm}\left(b_i + \sum_j W_{ij}\, h_j\right) \tag{5}$$

Generative training of a Deep Belief Network is done by training one RBM at a time. The goal of generative training is to determine the weights and biases that maximize the log-likelihood of the observed training data. For a fixed training data vector V, the gradient of the log-likelihood with respect to the weights $w_{ij}$ and biases $b_i$ and $c_j$ can be expressed as:

$$\frac{\partial \log P}{\partial w_{ij}} = \langle v_i h_j \rangle_{data} - \langle v_i h_j \rangle_{model} \tag{6}$$

$$\frac{\partial \log P}{\partial b_i} = \langle v_i \rangle_{data} - \langle v_i \rangle_{model} \tag{7}$$

$$\frac{\partial \log P}{\partial c_j} = \langle h_j \rangle_{data} - \langle h_j \rangle_{model} \tag{8}$$

The first term $\langle v_i h_j \rangle_{data}$ is the clamped expectation with V fixed and can be efficiently computed from the training data using equation (4). On the other hand, the second term $\langle v_i h_j \rangle_{model}$ is the expectation over the joint probability distribution defined by (1), i.e.

$$\langle v_i h_j \rangle_{model} = \frac{1}{Z} \sum_{\{v_k\}} \sum_{\{h_l\}} v_i\, h_j \exp\left(\sum_k b_k v_l + \sum_l c_l h_l + \sum_{kl} W_{kl} v_k h_l\right) \tag{9}$$

This term becomes intractable as the number of visible and hidden nodes increases.

To avoid computing the intractable term $\langle v_i h_j \rangle_{model}$, Contrastive Divergence (CD) based learning [10] does not directly use the formulas (6)-(8) to update the weights and biases. Instead, starting with a training vector $V^0$, the hidden layer values $H^0$ are computed by sampling from the conditional distribution (4). Then, a reconstruction of the visible layer $V^1$ is computed using (5), and then of the hidden layer $H^1$ using (4) again.

The weight updates are computed using

$$\Delta w_{ij} = \epsilon[\, \langle v_i h_j \rangle_{data} - \langle v_i h_j \rangle_{recon}\,] \tag{10}$$

(Similar formulas apply for the bias updates.)

Mathematically, the update formula (10) is an approximation which does not follow the gradient as in (6). Hinton et al have shown [11] that if the process of forward sampling and reconstruction were repeated $n$ times, then (10) would converge to the true gradient as $n \to \infty$, although in practice the updates are computed using just a single-step reconstruction. In practice this works well; however, CD may take many iterations to converge, due to the noise inherent in Gibbs sampling and slow mixing to the equilibrium probability distribution.

## 3 Quantum Sampling-Based Approach to Training Deep Neural Networks

In this section we describe our overall training approach, and give details on how a quantum annealing machine can be used to implement an RBM and generate samples to estimate the model expectations.

### 3.1 Overall training approach

Our overall approach to training Deep Neural Networks is depicted in Figure 2. The generative training utilizes a hybrid classical/quantum computing architecture. The outer loop of the training runs on a classical computer. Unlike conventional training approaches such as Contrastive Divergence, we compute our weight and bias updates based on the true gradient formulas (6)-(8). A quantum annealing machine, such as the D-Wave Two, is used to generate samples to estimate the "intractable" model expectation terms in these formulas.

#### 3.1.1 Generative Training (aka Pre-Training)

In the generative training, we train a DBN, one RBM at a time. For each RBM, we initialize the weights and biases to random values. We then calculate updates to the weights and biases using the gradient formulas (6)-(8):

$$w_{ij}^{(t+1)} = \alpha w_{ij}^{(t)} + \epsilon[\,\langle v_i h_j \rangle_{data} - \langle v_i h_j \rangle_{model}] \quad (11)$$
$$b_i^{(t+1)} = \alpha b_i^{(t)} + \epsilon[\,\langle v_i \rangle_{data} - \langle v_i \rangle_{model}] \quad (12)$$
$$c_j^{(t+1)} = \alpha c_j^{(t)} + \epsilon[\,\langle h_j \rangle_{data} - \langle h_j \rangle_{model}] \quad (13)$$

where $\alpha$ is the momentum and $\epsilon$ is the learning rate. The model expectations $\langle v_i h_j \rangle_{model}$, $\langle v_i \rangle_{model}$, and $\langle h_j \rangle_{model}$ are estimated using quantum sampling, following the method described in section 3.2 below.

If truth labels are available and the generative training is to be followed by discriminative training, we refer to the generative training as "pre-training".

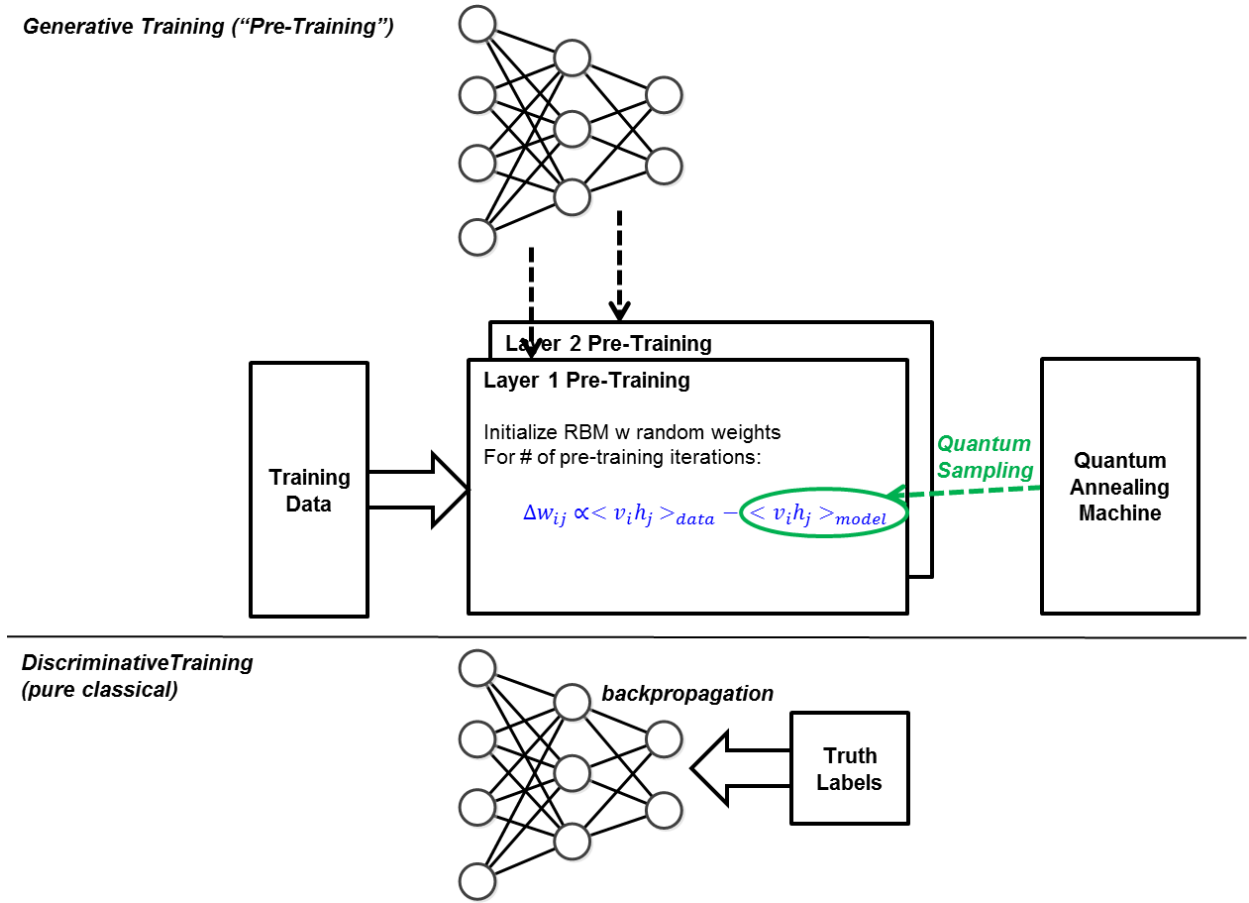

**Figure 2 Overall training approach including generative and discriminative training**

### 3.1.2 Discriminative Training

Discriminative training is done on a classical computer. The discriminative training is initialized by replacing the final pre-training weights in the last RBM layer with a least-squares regression from the last hidden layer to the truth labels, using a matrix pseudoinverse technique. While this is a somewhat less common initialization method than random weights, it is the method employed in the Deep Neural Network MATLAB toolbox [12] that we used to represent the conventional DBN training approach in our studies, hence we also use this method in our quantum sampling-based training approach, to facilitate an "apples to apples" comparison. Also, there have been some studies where backpropagation converged more quickly using this type of initialization.[13,14,15] Finally the standard backpropagation technique [16] is used to fine-tune the network weights and biases.

## 3.2 Quantum sampling approach

### 3.2.1 Quantum annealing and Boltzmann sampling

There is a natural correspondence between the RBM energy functional (2) and the Hamiltonians used in quantum annealing. In quantum annealing, the Hamiltonian evolves over time from an initial Hamiltonian $\mathcal{H}_i$ to a final Hamiltonian $\mathcal{H}_f$:

$$\mathcal{H}(t) = (1 - s(t))\mathcal{H}_i + s(t)\mathcal{H}_f \qquad 0 \leq t \leq T \qquad (14)$$

where $s(t)$ increases from 0 to 1 as $t$ goes from 0 to $T$. In an ideal adiabatic evolution, if the system starts in the ground state of $\mathcal{H}_i$ and the evolution proceeds slowly enough to satisfy the conditions of the adiabatic theorem, then the system will end up in the ground state of $\mathcal{H}_f$.

However in an actual hardware implementation of quantum annealing, such as the D-Wave machine, there is inevitably some interaction between the qubits and their environment, which leads to a nonzero probability that the evolution will end up in an excited state. While this non-ideality of actual quantum annealing hardware may be an undesirable feature when it comes to optimization, we may be able to take advantage of this non-ideality to utilize the quantum annealer as a sampling engine.

We conjecture that the distribution of excited states can be modeled, at least approximately, as a Boltzmann distribution:

$$P(x) = \frac{1}{Z} \exp\left(-\beta_{eff} \mathcal{H}_f(x)\right) \qquad (15)$$

Since the RBM joint probability distribution (1) also has this form, our basic plan is as follows:

- Use the RBM energy functional (2) as the final Hamiltonian $\mathcal{H}_f$
- Run quantum annealing N times and take the sample average

$$\overline{v_i h_j} = \frac{1}{N} \sum_{n=1}^{N} v_i^{(n)} h_j^{(n)} \qquad (16)$$

  as our estimate of the model expectation $\langle v_i h_j \rangle_{model}$ (and similarly for the expectations $\langle v_i \rangle_{model}$ and $\langle h_j \rangle_{model}$).

However, this plan needs to be refined to address several potential obstacles:

- **Limited qubit connectivity.** The RBM energy functional (2) reflects the bipartite graph connectivity between the visible and hidden nodes. On the other hand, practical implementations of quantum annealing, such as the D-Wave machine, may have sparser connectivity between qubits such that the RBM bipartite graph cannot be mapped directly onto the hardware graph.

Figure 3(a) shows the qubit connectivity in a 4-unit cell subset of a D-Wave chip. Each 8-qubit unit cell consists of 4 vertical qubits (shown in blue) and 4 horizontal qubits (shown in green). Each circle represents a Josephson junction where a pair of qubits can be coupled. Each interior qubit is coupled to 6 other qubits, 4 within the same unit cell and 2 in the adjacent unit cells. The resulting topology, shown in Figure 3(b), is known as a "Chimera" graph, consisting of a square lattice of $K_{4,4}$ bipartite subgraphs.

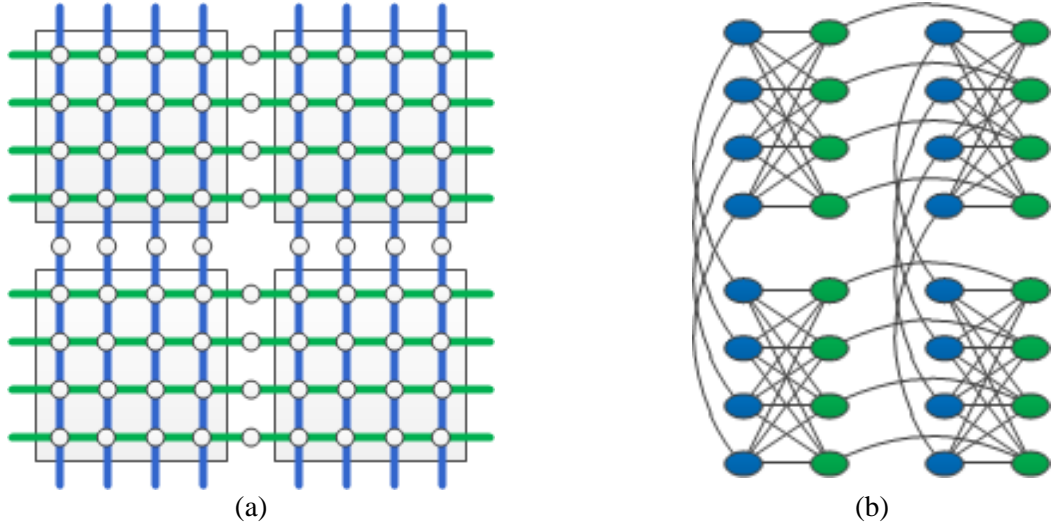

(a)            (b)

**Figure 3 Two representations of D-Wave qubit connectivity: (a) Chip layout; (b) "Chimera" graph**

Our approach overcomes this limited qubit connectivity by mapping each RBM visible and hidden node to a chain of qubits, as described below.

- **Faulty qubits.** Due to fabrication and calibration issues, a small number of the qubits on the D-Wave chip are not usable (during this study, there were 8 faulty qubits on the 512-qubit "Vesuvius" chip operated by Lockheed Martin). Again, this complicates the process of mapping the RBM bipartite graph onto the hardware graph. Our approach of mapping each RBM visible and hidden node to a chain of qubits, also compensates for these faulty qubits.
- **Parameter setting noise (aka intrinsic control error).** The D-Wave machine exhibits intrinsic control errors such that the effective values of the final Hamiltonian coefficients may deviate slightly from their programmed values. Currently these coefficients can be programmed with 4 bits of precision. Our approach uses *gauge transformations* (defined below) to partially mitigate the effects of these intrinsic control errors.
- **The parameter $\beta_{eff}$ in (15) is undetermined.** Assuming that the distribution of samples from the quantum annealing follows a Boltzmann distribution, there is an overall scale factor $\beta_{eff}$ between the RBM energy functional (2) and the final Hamiltonian $\mathcal{H}_f$ that is undetermined. Note that since the final Hamiltonian is programmed on the D-Wave machine using dimensionless coefficients, the parameter $\beta_{eff}$ is not the usual $\beta = 1/kT$. It can be viewed as an empirical parameter that depends on the operating temperature of the hardware, the energy scale of the superconducting flux qubit system, the coupling of the qubits to the environment, and the quantum annealing time evolution. Our approach includes a process for selecting a value for $\beta_{eff}$ based on the size (number of visible and hidden nodes) of the RBM.

With the refinements added to overcome these obstacles, we have been able to achieve good practical results in training Deep Neural Networks, as described in section 4 below.

### 3.2.2 Mapping RBMs onto the quantum annealing hardware graph

Assume that our RBM has energy functional (2). If we define the binary vector $x$ to be the concatenation of the visible node vector $v$ and the hidden node vector $h$, then we can write (2) in the form

$$E = \beta_{eff} \, x^T Q x \tag{17}$$

where $Q$ is the $(n+m) \times (n+m)$ matrix

$$Q = \frac{1}{\beta_{eff}} \begin{bmatrix} B & W \\ 0 & C \end{bmatrix} \tag{18}$$

where B and C are diagonal matrices containing the biases $\{b_i\}$ and $\{c_j\}$ respectively. The scale factor $\beta_{eff}$ is at this point still undetermined.

On a quantum annealing machine like the D-Wave, it is often more convenient to work with "spin" variables that take values $\pm 1$, rather than binary 0/1 variables. The transformation

$$x \to S = 2x - 1 \tag{19}$$

induces a transformation from the matrix $Q$ to an ***Ising model (H, J)***

$$E' = -\sum_{i=1}^{n} H_i S_i - \sum_{j=n+1}^{n+m} H_j S_j - \sum_{i=1}^{n} \sum_{j=n+1}^{n+m} J_{ij} S_i S_j \tag{20}$$

where the first n spin variables correspond to the visible nodes, and the next m spin variables correspond to the hidden nodes.

Next, we embed this Ising model onto the D-Wave chip by mapping each visible node to a chain of vertical qubits, and each hidden node to a chain of horizontal qubits, as in Figure 4 below. The embedded Ising model has the form

$$E'' = E' - J_{FM} \sum_{i=1}^{n} \sum_{(k,l) \in emb\{i\}} S_i^{(k)} S_i^{(l)} - J_{FM} \sum_{j=n+1}^{n+m} \sum_{(k,l) \in emb\{j\}} S_j^{(k)} S_j^{(l)} \tag{21}$$

where the extra terms have a ferromagnetic coupling $J_{FM}>0$ to enforce that qubits that are chained together should agree with each other. In the example shown in Figure 4, each node in the 8x8 RBM maps to 2 physical qubits, which are chained together with ferromagnetic couplings, shown as black circles. Note that there are 64 intra-unit-cell couplers, shown as red circles, each of which can be independently programmed, allowing all the connection weights in the RBM to be modeled.

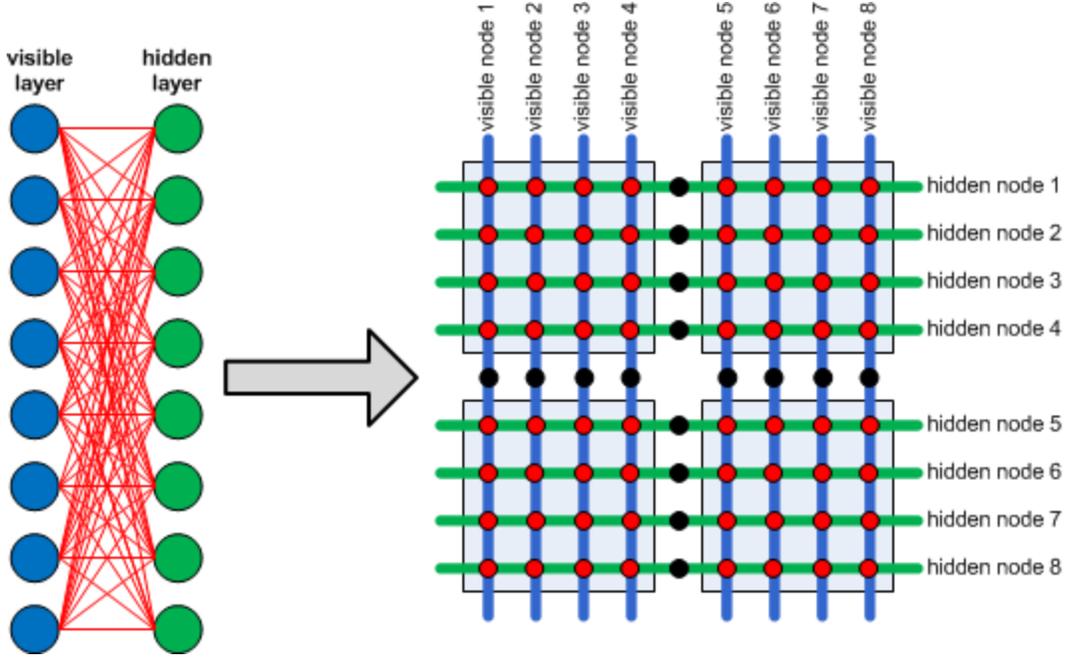

**Figure 4 Embedding of RBM visible and hidden nodes onto D-Wave chip layout**

On an ideal 512-qubit Vesuvius chip, this mapping enables us to realize a full 32x32 bipartite graph. On the real Vesuvius chip operated by Lockheed Martin, which has 8 faulty qubits, 32 of the 1024 visible-to-hidden node connections cannot be implemented. In our approach, we assume those connection weights $W_{ij}$ must be zero, so our trained model will not use those weights. In our experiments we have found that the absence of a small number of connections does not affect the performance of the resulting network.

While the ferromagnetic coupling $J_{FM}$ is intended to discourage configurations where the qubits in a chain disagree with each other, there is a small probability that this will occur. This is somewhat more likely if a chain includes a faulty qubit, so that the two segments of the chain are not directly connected. Our approach includes a tunable *voting threshold* parameter to control how strictly the chain constraints are enforced. If the voting threshold is set to 1.0 (strict enforcement), then any samples that violate the chain constraints will be discarded before the sample averages are computed. The theoretical Boltzmann distribution for the embedded Ising model $E''$ should have a negligible contribution from these configurations due to the energy penalty from $J_{FM}$. The remaining configurations, the ones that satisfy all the chain constraints, should all yield the same energy for the $J_{FM}$ terms, so this factor will cancel out of the numerator and denominator (the partition function *Z*) in the expectation. Thus, to a good approximation,

$$\langle v_i h_j \rangle_{model}^{E''} \cong \langle v_i h_j \rangle_{model}^{E} \qquad (22)$$

so that sample averages from the embedded Ising model should approximate the model expectations from the original RBM joint distribution.

If the voting threshold is set to a value *r* with $0.5 \leq r < 1$, then if a fraction greater than *r* of the qubits in a chain agree with each other, then the configuration will be accepted, and the majority value will be used as the value for that chain. Theoretically, using a threshold $r < 1$ gives slightly less fidelity between sampling from the embedded Ising model and the original RBM joint distribution; on the other hand, it reduces sampling noise by allowing more samples to be used out of a given number of annealing runs, than with the strict enforcement $r = 1$. In practice, we have found that majority voting tends to result in slightly more accurate estimates for the model expectations than the strict enforcement.

### 3.2.3 Correcting for parameter noise

The D-Wave machine exhibits intrinsic control errors such that the effective values of the final Hamiltonian coefficients may deviate slightly from their programmed values. The Vesuvius V6 processor used in this study provides 4 bits of precision in setting these coefficients. In some cases, the errors are systematic in nature. In particular, when a coupling $J_{ij}$ is programmed on the machine, there is a small "leakage" that effectively adds a bias of roughly 3%, of the same sign as $J_{ij}$, to the individual qubits $i$ and $j$. When confined to a single coupler, this effect is not noticeable. However, when multiple qubits are chained together using ferromagnetic couplings, which was part of our approach to overcome the limited qubit connectivity, this can skew the statistics.

To mitigate this D-Wave coupler/bias "leakage" effect, we use *gauge transformations*.[17] One of the gauge transformations we use is shown in Figure 5.

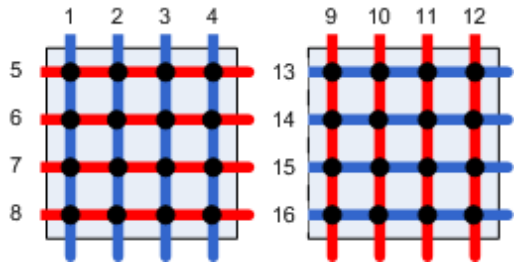

**Figure 5 Gauge transformations are used to partially mitigate intrinsic control errors**

In the figure, the red qubits are "flipped" by the gauge transformation while the blue qubits are unchanged. In the first unit cell, all the horizontal qubits are flipped while the vertical qubits are unchanged. In the next unit cell, all the vertical qubits are flipped while the horizontal qubits are unchanged. Because of the way this pattern alternates from one unit cell to the next, we call this the "basket weave" or "checkerboard" gauge transformation.

A gauge transformation on the qubits induces a transformation on the Ising model coefficients $H_i$ and $J_{ij}$. One interesting property of the gauge transformation in Figure 5 is that it transforms all ferromagnetic couplings into antiferromagnetic couplings, and vice versa. This is especially useful in mitigating the coupler/bias leakage effect in our chained qubits.

Similarly, if individual qubits are biased in one direction or the other, this can be mitigated by a gauge transformation that flips all the spins.

Thus in our approach, when we sample from the embedded Ising model $E''$, we average over 4 gauges:

- I – the identity transformation (no spins flipped)

- G – the gauge transformation shown in Figure 5 that flips half the spins on the chip
- -G – the gauge transformation that flips the other half of the spins on the chip
- -I – the transformation that flips all the spins

### 3.2.4 Selecting a value for $\beta_{eff}$

As noted above, if we assume that the distribution of samples from quantum annealing can be modeled as a Boltzmann distribution as in (15), there is an overall scale factor $\beta_{eff}$ between the RBM energy functional (2) and the final Hamiltonian $\mathcal{H}_f$ that is undetermined.

In our approach we select a value for $\beta_{eff}$ by constructing an RBM of a particular size; mapping that RBM to a final Hamiltonian assuming a particular value of $\beta_{eff}$; running quantum annealing; computing the model expectations using the quantum samples; and then comparing the resulting expectations to the "correct" values. We repeat this process for different choices of $\beta_{eff}$ and select the value of $\beta_{eff}$ that gives us the best fit.

For RBMs up to size 16x16, we generated RBMs with random (W,b,c) and calculated the model expectations $\langle v_i h_j \rangle_{model}$ exactly using brute force (evaluating all 2^(n+m) possible configurations. For RBMs of size 32x32, we used Markov Chain Monte Carlo sampling to estimate the model expectations.

To compute the expectations using quantum sampling, for each candidate value of $\beta_{eff}$ we ran 40 iterations, where each iteration consisted of 4 solver calls, one for each of the 4 gauges defined in section 3.2.3, and each solver call to the D-Wave made 1000 annealing runs, each with an annealing time of 20 $\mu$s. We measured the error of the quantum sampling-based expectation from the "correct" expectation using the $L_1$-norm, $\left\| \langle v_i h_j \rangle_{quantum} - \langle v_i h_j \rangle_{correct} \right\|_1$.

We found that the optimal setting for $\beta_{eff}$ is size dependent. For smaller size RBMs (e.g. 5x5 and 8x8), we got the best fit to the exact expectations using $\beta_{eff}$=4.5. However as we increased the size (12x12, 14x14, 16x16), we got a better fit using $\beta_{eff}$=3. Finally for 32x32, we got the best fit using $\beta_{eff}$ =2.

Even with the optimal settings for $\beta_{eff}$, the estimates of the model expectations will still have some error, but compared to the noise associated with Gibbs sampling in Contrastive Divergence, this may be sufficient to estimate the gradient in (6)-(8).

## 4 Experimental Results

We tested our approach using the MNIST data set of handwritten digits [18], which is widely used as a benchmark for machine learning algorithms. The MNIST data set contains 60,000 training and 10,000 test set images with truth labels. Each image consists of 784 greyscale pixels (28x28) representing handwritten digits from 0-9.

Since the image size of 784 pixels exceeds the maximum input layer size of 32 nodes that can currently be realized on the D-Wave machine, we scaled each image down as follows:

- The 2 pixels on the boundary were discarded, leaving a 24x24 image.
- We compute the average pixel value over each 4x4 block, resulting in a coarse-grained 6x6 image.
- We discarded the 4 corners, resulting in 32 super-pixels.

We refer to the resulting images as the Coarse-Grained MNIST (CG-MNIST) data set.

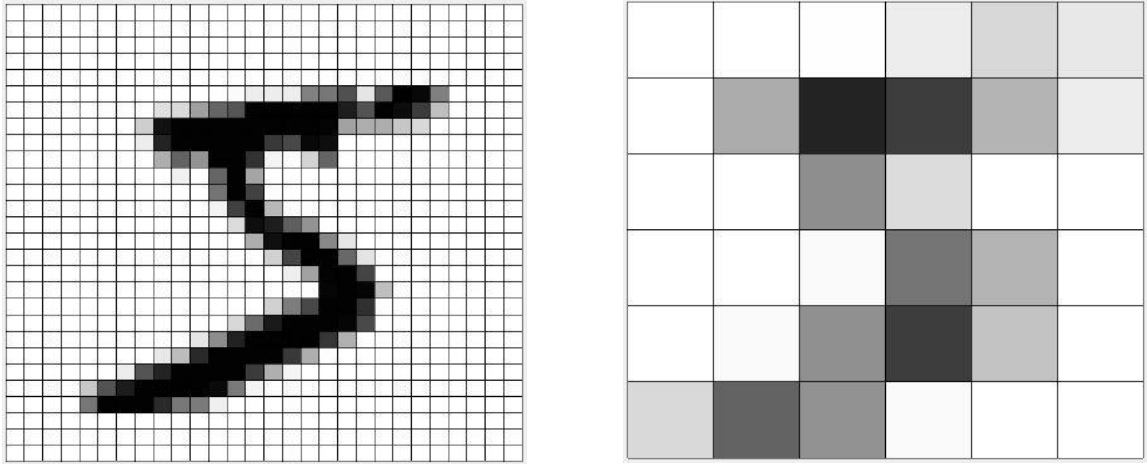

**Figure 6 Example of original (28x28) and coarse-grained (6x6) MNIST images of handwritten "5"**

Figure 6 shows an example of the coarse-graining process for the first image in the MNIST training set, for which the truth label is the digit "5". (The right hand image still has 36 super-pixels, but the 4 corners are not used.) As can be seen, the coarse-grained MNIST data set is actually a more difficult classification problem due to the ambiguity of the images. While state-of-the-art methods have been able to achieve over 99.7% test set accuracy on the real MNIST data set [18], it would be unreasonable to expect this level of accuracy on our coarse-grained MNIST data set.

For comparison purposes, we used as our classical baseline, the Deep Neural Network MATLAB toolbox by M. Tanaka.[12] This toolbox implements training for Deep Belief Nets and Restricted Boltzmann Machines including Contrastive Divergence.

After generating coarse-grained versions of the MNIST training and test data sets, we compared the classical training approach (using the Deep Neural Network toolbox) to the quantum sampling-based training approach (described in section 3). For brevity, we will refer to the two approaches below as the "classical" and "quantum" approaches respectively. All classical and quantum DBNs were constructed with 32 input nodes, 2 hidden layers with 32 nodes each, and an output layer with 10 nodes (one for each possible digit). The mappings of the 32x32 RBMs use the entire D-Wave Two chip, which includes 8 faulty qubits.

**Generative Training (Pre-Training).** For both the classical and quantum pre-training, we used identical settings for the common experimental parameters. We divided the MNIST training set into 10 sets of 6,000 images each. A classical and quantum DBN of size 32/32/32/10 DBN were trained on each of the 10 6,000-image sets for N pre-training iterations, where N ranged from 1 to 50. This means for each N there were 10 networks trained, and so each data point in Figure 7 through Figure 10 below represents an average over 10 trials. We used a learning rate $\epsilon = 0.1$. Momentum was set to $\alpha = 0.5$ for the first 5 training iterations and $\alpha = 0.9$ afterwards. Mini-batching was not used.

For the quantum networks, on each pre-training iteration we obtained 100 samples from the D-Wave in each of the 4 gauges (total 400 samples). We used a voting threshold $r = 0.5$. $\beta_{eff}$ was set to 2 for the first two RBMs of 32/32 size and $\beta_{eff}$ was set to 3 for the final RBM of size 32/1 (see section 3.2.4 for more details).

**Discriminative Training.** Discriminative training was performed identically for both classical and quantum. After completing N pre-training iterations, we applied the truth labels and set the last RBM layer coefficients using linear regression. This was followed by 1000 iterations of backpropagation using mini-batches of size 100.

A subset of the results are shown in Figure 7 through Figure 10, comparing the classical and quantum results after 100, 200, 400, and 800 iterations of backpropagation respectively. In each figure, the left (red) plot is the classical training, and the right (blue) plot is the quantum sampling-based training. Along the horizontal axis are the number of pre-training iterations. The vertical axis is accuracy, where the training set accuracy (averaged over 10 trials) is plotted as a dashed line, and the test set accuracy (averaged over 10 trials) is plotted as a solid line. Error bars indicate $\pm 1$ standard deviation for the 10 trials.

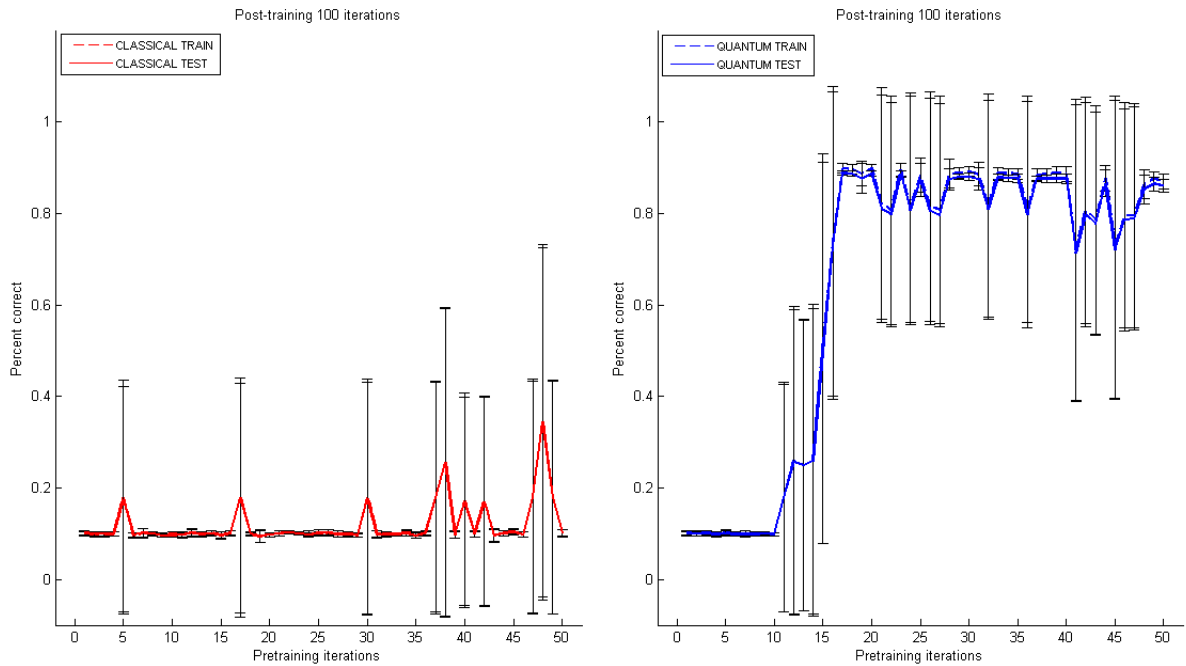

**Figure 7 CG-MNIST: Classical vs. quantum sampling-based training after 100 post-iterations**

Application of Quantum Annealing to Training of Deep Neural Networks

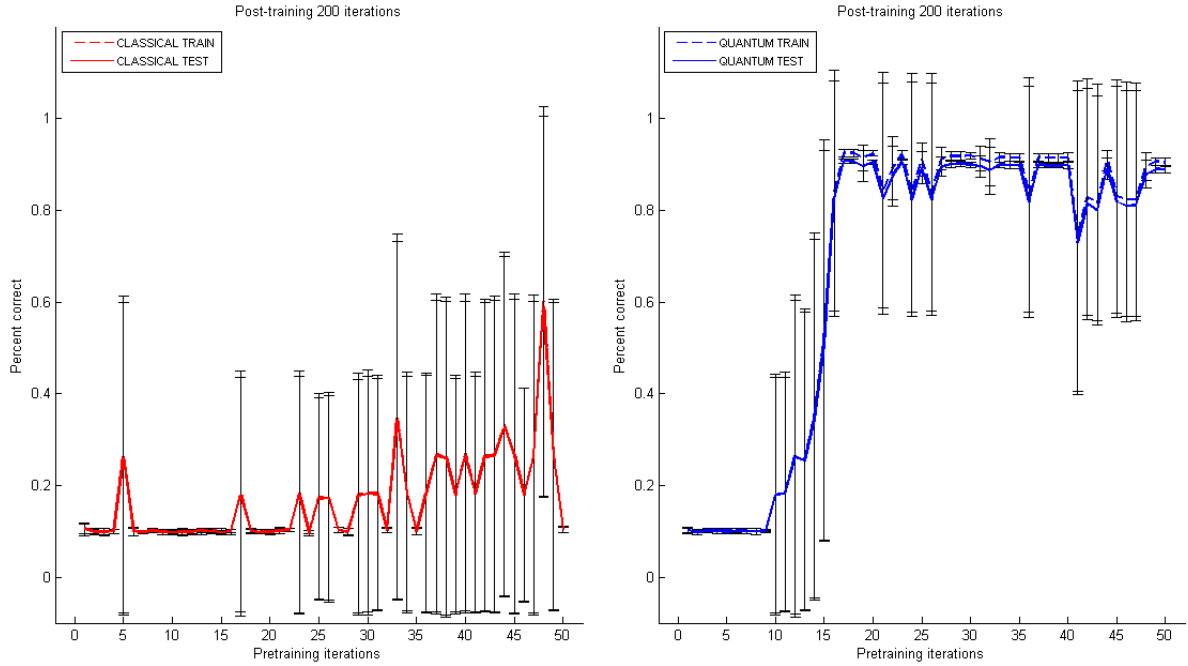

**Figure 8 CG-MNIST: Classical vs. quantum sampling-based training after 200 post-iterations**

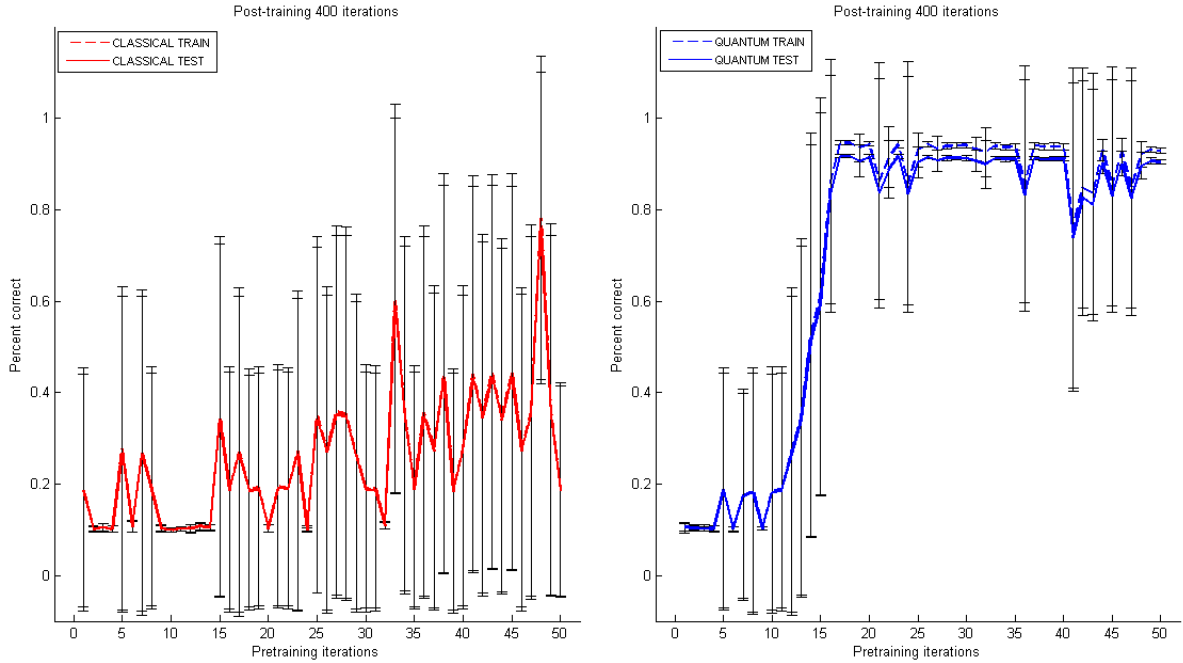

**Figure 9 CG-MNIST: Classical vs. quantum sampling-based training after 400 post-iterations**

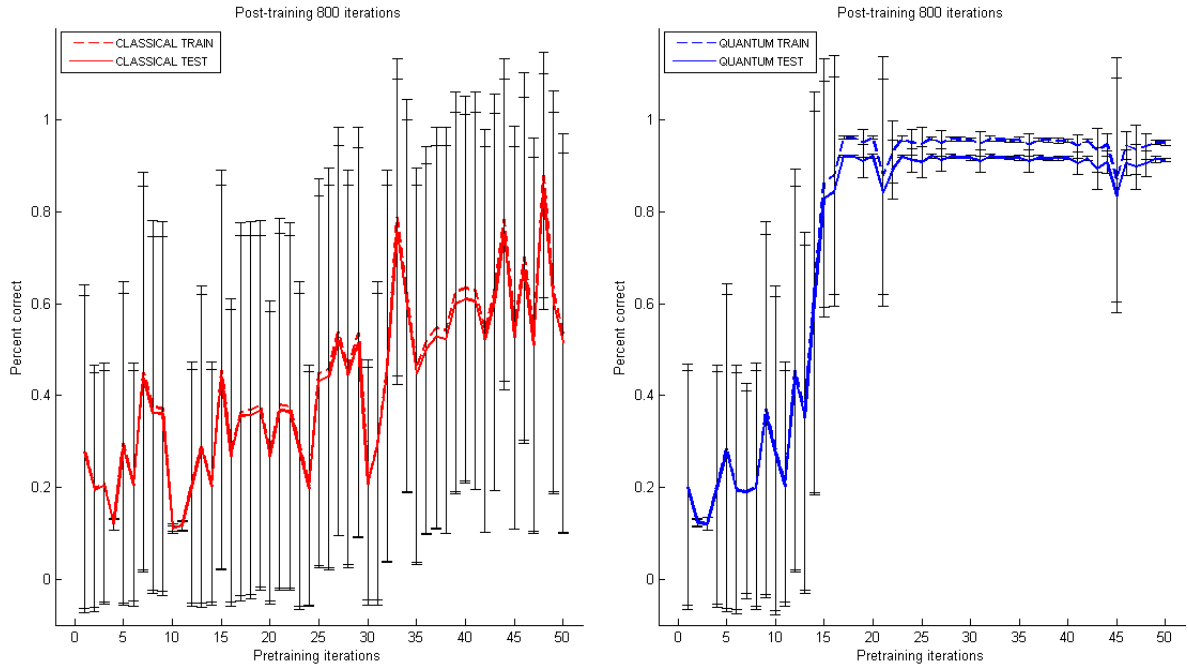

**Figure 10 CG-MNIST: Classical vs. quantum sampling-based training after 800 post-iterations**

While both the classical and quantum networks improve in accuracy as the number of pre-training and backprop iterations increase, the quantum sampling-based training appears to require significantly fewer iterations of both pre-training and backprop to reach a given level of accuracy. For example, comparing Figure 7 and Figure 10, quantum networks trained with only 20 pre-training and 100 backprop iterations were more accurate than classical networks trained for 50 pre-training and 800 backprop iterations. In addition, we see much smaller variation in accuracy among the quantum networks that were trained for the same number of pre-training and backprop iterations, than with the classical networks; this difference becomes especially pronounced in Figure 10 (800 backprop iterations) for quantum networks that have had more than 20 pre-training iterations.

## 5 Discussion

The results we have obtained so far suggest that the quantum sampling-based training approach could significantly reduce the amount of generative training, as well as discriminative training, needed to achieve the same level of post-training accuracy.

The reduction in generative training iterations lends support to the hypothesis that sampling from a quantum annealing device will mix faster than classical Gibbs sampling, due to quantum superposition and tunneling. However, the results we have obtained so far are purely empirical. While the curves in Figure 7 through Figure 10 appear to show qualitatively different "classical" and "quantum" behaviors, further theoretical and experimental investigations are needed to understand whether this can truly be attributed to quantum effects. Previous attempts to compare the performance of classical and quantum devices have illustrated that defining and detecting quantum "speedup" can be fraught with difficulties.[19] Similarly in this case, it is conceivable that the behavior on the right hand side of these figures could be reproduced by a cleverly-designed classical algorithm.

Since real hardware implementations of quantum annealing will have imperfections such as processor noise and possibly defective qubits, it is also encouraging that our results show that the method is robust enough to produce trained networks with high accuracy in spite of these imperfections. For example, to handle the coarse-grained MNIST data set, our mapping of the 32x32 RBM onto the chip is missing 32 out of 1024 possible connections, but the quantum sampling-based training works just as well, and in fact better, than the classical training.

As D-Wave produces larger chips and increases the connectivity between qubits, we expect that the size of the RBM layers that can be trained by this method will increase. For example, with the current 8-qubit unit cell architecture, a 64x64 RBM could be mapped onto a square 2048-qubit chip using the mapping defined in section 3.2.2. Chip designs with larger unit cells would accelerate this scaling.

# 6  Conclusions

We investigated an approach for training Deep Neural Networks, based on sampling from a quantum annealing machine such as the D-Wave device. Our approach for mapping RBMs onto the quantum hardware overcomes the limitations of physical qubit connectivity and faulty qubits. We tested our approach on a scaled-down version of the MNIST data set using coarse-grained images, and found that the quantum sampling-based training approach can lead to comparable post-training accuracy to classical Contrastive Divergence training with fewer iterations of generative and discriminative training.

Further investigation is needed to determine whether similar improvements can be achieved for other data sets, and to what extent these improvements can be attributed to quantum effects. As larger quantum annealing machines become available in the future, further investigation is needed to characterize how the performance of the quantum sampling-based training approach scales relative to conventional training as the size of the RBM increases.

In addition, the idea of using a quantum annealing machine to do sampling or inference, as opposed to optimization, could lead to new applications of quantum annealing in addition to Deep Learning. Exact inference on probabilistic graphical models becomes intractable as the number of nodes increases, except when the graph is a tree, and approximate inference methods such as loopy belief propagation are not always guaranteed to converge.[9] Similarly, sampling methods such as Markov Chain Monte Carlo, can have very long mixing times. Quantum sampling, using a quantum annealing machine to draw representative samples from a Boltzmann distribution, could potentially provide an alternative to conventional techniques for sampling and inference in some cases.

**Acknowledgments**

This work was supported by Internal Research & Development funding from Lockheed Martin.

Thanks to Dan Davenport and Kristen Pudenz for useful discussions. Thanks to Paul Alsing and Daniel Lidar for helpful feedback on earlier versions of this manuscript.

The LOCKHEED MARTIN trademark used throughout is a registered trademark in the U.S. Patent and Trademark Office owned by Lockheed Martin Corporation.